\documentclass[10pt]{jfm}
\usepackage{physics}
\usepackage[usenames, dvipsnames]{color}
\usepackage{ar}
\usepackage{amsmath,color,graphicx,amssymb}
\usepackage[font=small]{caption}
\usepackage{amsbsy}
\usepackage{latexsym}
\usepackage[normalem]{ulem}
\usepackage[english]{babel}
\usepackage{ifsym}
\usepackage{bbding}
\usepackage{wasysym}
\usepackage{latexsym}
\usepackage{epsfig} 
\usepackage{epsf,psfig}
\usepackage{epstopdf}
\usepackage{multirow}
\usepackage{array}
\usepackage{float}
\usepackage[english]{babel}
\usepackage{verbatim}
\usepackage{soul}
\usepackage[toc,page]{appendix}
\usepackage{setspace}
\usepackage{threeparttable}
\usepackage{booktabs,array}
\usepackage{color}
\usepackage{mathtools}
\usepackage{siunitx}
\usepackage{setspace}
\usepackage[numbers]{natbib}
\usepackage{authblk}
\usepackage{gensymb}
\usepackage[section]{placeins}
\usepackage{caption}
\definecolor{lightblue}{RGB}{48, 89, 155}
\definecolor{black}{RGB}{0, 0, 0}

\usepackage{float}
\def\bibfont{\footnotesize}
\usepackage{trimclip}
\def\vk{von K\'{a}rm\'{a}n }

\begin{document}

\newtheorem{lemma}{Lemma}
\newtheorem{corollary}{Corollary}

\shorttitle{Fundamental forces behind equilibrium altitudes of near-ground swimmers } 
\shortauthor{Tianjun Han et al.} 

\title{Revealing the mechanism and scaling laws behind equilibrium altitudes of near-ground pitching hydrofoils}


\author
 {
Tianjun Han\aff{1}
  \corresp{\email{tih216@lehigh.edu}},
  Qiang Zhong\aff{2,3},
  Amin Mivehchi\aff{1},
  Daniel B. Quinn\aff{2}
  and Keith W. Moored\aff{1}
  }

\affiliation
{
\aff{1}
Department of Mechanical Engineering and Mechanics, Lehigh University, Bethlehem, PA 18015, USA
\aff{2}
Department of Mechanical and Aerospace Engineering, University of Virginia, Charlottesville, VA 22904, USA
\aff{3}
Department of Mechanical Engineering, Iowa State University, Ames, IA 50011, USA
}

\graphicspath{{Figures/}}
\maketitle

\begin{abstract}
A classic lift decomposition \citep{Sears1938} is conducted on potential flow simulations of a near-ground pitching hydrofoil.  It is discovered that previously observed stable and unstable equilibrium altitudes are generated by a balance between positive wake-induced lift and negative quasi-steady lift while the added mass lift doesn't play a role. Using both simulations and experiments, detailed analyses of each lift component's near-ground behavior provide further physical insights. When applied to three-dimensional pitching hydrofoils the lift decomposition reveals that the disappearance of equilibrium altitudes for $\AR<1.5$ occurs due to the magnitude of the quasi-steady lift outweighing the magnitude of the wake-induced lift at all ground distances. Scaling laws for the quasi-steady lift, wake-induced lift and the stable equilibrium altitude are discovered. A simple scaling law for the lift of a steady foil in ground effect is derived.  This scaling shows that both circulation enhancement and the velocity induced at a foil’s leading edge by the bound vortex of its ground image foil are the essential physics to understand steady ground effect. The scaling laws for unsteady pitching foils can predict the equilibrium altitude to within $20\%$ of its value when $St < 0.45$.  For $St \ge 0.45$ there is a wake instability effect, not accounted for in the scaling relations, that significantly alters the wake-induced lift. These results not only provide key physical insights and scaling laws for steady and unsteady ground effect, but also for two schooling hydrofoils in a side-by-side formation with an out-of-phase synchronization.
\end{abstract}

\section{Introduction}
Many animals improve their locomotion energetics or increase their range by swimming or flying near the seafloor or water surface, thereby taking advantage of ground effect \citep{Webb1993, HAINSWORTH1988, Baudinette1974, Blake1979, Park2010}. When wings/fins move steadily near a boundary, such as those of gliding animals or fixed-wing aircraft, it is well-known that a \textit{steady ground effect} increases their lift and reduces their induced drag \citep{TremblayDionne2018, He2015, Lee2011, Qu2015, Yang2010, Yang2009, Boschetti2010, Widnall1970, Baddoo2020}. However, when swimmers or flyers oscillate their appendages near a boundary, like flatfish swimming near the ocean floor, they experience an \textit{unsteady ground effect}.

Unsteady ground effect produces a host of hydrodynamic phenomena. For example, foils oscillating near a boundary experience a thrust or cruising speed enhancement with no penalty to their propulsive efficiency \citep{Wu2014, Park2017, Dai2016, Fernandez-Prats2015, Quinn2014b}, and vortex wakes are deflected near boundaries due to their image vortices \citep{Quinn2014c, Kurt2019, Zhong2019}.  Moreover, stable equilibrium altitudes arise where an increase/decrease in a swimmer's altitude causes a negative/positive lift force, pulling the swimmer back to a stable altitude \citep{Quinn2014c,Mivehchi2016}. \citet{Kurt2019} showed that pitching foils swimming freely in the cross-stream direction do indeed settle into a stable near-ground equilibrium altitude under no control. By showing good agreement between the equilibrium altitudes calculated through inviscid potential flow simulations and those measured in viscous water channel experiments, they determined that stable equilibrium altitudes are an inviscid phenomenon \citep{Kurt2019}. \citet{Zhong2019} broadened our understanding of these equilibrium altitudes by studying hydrofoils of varying aspect ratio. They discovered that near-ground equilibrium altitudes can be stable, unstable, or absent depending upon the aspect ratio and Strouhal number.

Now, it's essential to understand the underlying physics that give rise to equilibrium altitudes.  In this study, we uncover the fundamental forces that balance to generate equilibrium altitudes. Following the classic force decomposition of \citet{Sears1938} and \citet{Mccune2014}, we advance our understanding in two important ways: (1) we reveal the mechanism behind equilibrium altitudes and (2) we leverage this understanding to develop scaling laws to predict equilibrium altitudes from only the input kinematics.  Previously, \citet{Kurt2019} proposed that stable equilibrium altitudes are generated by a balance of the added mass lift, wake-induced lift, and quasi-steady lift. However, their force decomposition approach implicitly assumed that the local effective flow velocity acting on a foil was unaltered due to the presence of the ground.  In this study, we no longer make this assumption and, in fact, we determine that this effect is integral to our understanding of the mechanism behind equilibrium altitudes.  We will show that, upon deeper investigation, equilibrium altitudes exist due to a balance between the positive wake-induced lift and negative quasi-steady lift with added mass lift playing precisely zero role --- effectively a scallop theorem result \citep{purcell1977} for the time-averaged lift. In this way, our new understanding of the stable equilibrium altitude mechanism supersedes that which was proposed in \citet{Kurt2019}.

\section{Methods} \label{s:methods}

\subsection{Problem formulation}
\begin{figure}
    \centering
    \includegraphics[width=0.88\textwidth]{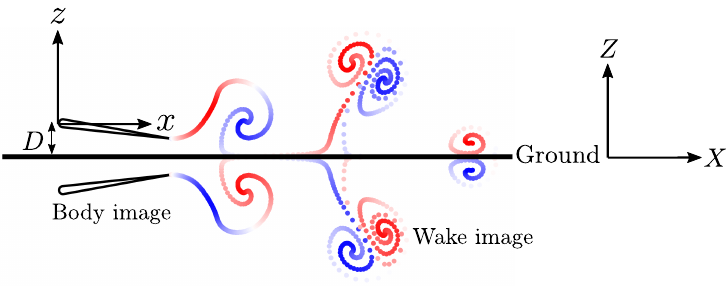}\vspace{0.in}
    \vspace{-0.1in}
    \caption{Illustration of a hydrofoil pitching near the ground. An inertial frame of reference fixed to the undisturbed fluid is denoted by $(X,Z)$ while a body-fixed frame of reference is denoted by $(x,z)$. An image system enforces a no-flux boundary condition at the ground plane ($Z = 0$).} 
    \label{fig:foilandwake}
\end{figure}
In this study, a simple model of a pitching hydrofoil is used to represent a swimmer propelling itself in unsteady ground effect (Figure \ref{fig:foilandwake}). The foil is a $7\%$ thick tear-drop shaped cross-section \citep{Godoy-Diana2008, Quinn2014c, Kurt2019} with a chord length of $c=0.095\,$m. The foil undergoes sinusoidal pitching about its leading edge,
\begin{align} 
\theta(t)=\theta_{0}\sin(2\pi f t ),
\end{align}
\noindent where $\theta(t)$ is the instantaneous pitching angle, $\theta_{0}$ is the pitching amplitude, $f$ is the pitching frequency, and $t$ is time. The pitching angle is considered positive for clockwise pitching rotations about the leading edge. The trailing edge peak-to-peak amplitude $A$ is defined by
\begin{align}  
A = 2c\sin\theta_{0},
\end{align}
\noindent and its dimensionless counterpart is $A^* = A/c$. The reduced frequency and the Strouhal number are defined as
\begin{align}  
k=\frac{fc}{U_{\infty}} \qquad St = \frac{fA}{U_{\infty}}
\end{align}
\noindent where $U_\infty$ is the swimming speed of the foil. The ground proximity, $D$, is considered as the distance from the foil's leading edge to the ground plane (Figure \ref{fig:foilandwake}) and its dimensionless form is $D^* = D/c$. Finally, the lift force $L$ is normalized by the dynamic pressure to give lift coefficient,
\begin{align} 
C_{L} = \frac{L}{1/2\rho cs U_{\infty}^2},
\end{align}
where $\rho$ is the fluid density while the span length $s$ is set to unity for two-dimensional simulations, and $s = \AR c$ for three-dimensional simulations with $\AR$ being the aspect ratio. And its time-averaged form is given as
\begin{align} 
\overline{C}_{L} = \frac{\overline{L}}{1/2\rho cs U_{\infty}^2},
\end{align}
where $\overline{L}$ is the time-averaged lift.


\subsection{Numerical methods} \label{sec:num_methods}
The simulations model a potential flow that is assumed to be irrotational, incompressible, and inviscid. This flow is governed by Laplace's equation, $\nabla^{2}\Phi^{*}=0$, where $\Phi^{*}$ is the perturbation velocity potential in an inertial frame fixed to the undisturbed fluid. A no-flux boundary condition, $\nabla \Phi^*\,\boldsymbol{\cdot\, n}=0$ where $\boldsymbol{n}$ is the surface normal vector, must be satisfied on the body surface and the ground plane at $Z=0$, and a far-field boundary condition that flow perturbations decay far away from the body and its wake surfaces must be satisfied. 

Following previous studies \citep{Katz1985, Quinn2014c, Moored2018, Kurt2019, Zhong2019, Mivehchi2021}, this potential flow problem is solved numerically using an unsteady boundary element method with the method of images used to enforce the ground plane boundary condition.  
In this method, the foil, its wake surfaces, and their images are discretized into a finite number of boundary elements. 
The general solution to the potential flow problem is then reduced to finding a distribution of constant-strength doublet and source elements that satisfy the body and ground plane boundary conditions. 
The boundary elements also implicitly satisfy the far-field boundary condition.
The body boundary condition is imposed using a Dirichlet condition at collocation points centered on each body element, but pushed within the body along the element's surface normal. 
Also, an explicit Kutta condition is applied where the vorticity at the trailing edge is zero by tuning the strength of a wake boundary element at the trailing edge. 
At each time step, the trailing edge element is shed with a strength that satisfies Kelvin's circulation theorem and it remains constant thereafter. 
A wake roll-up algorithm is implemented to advect the wake elements with the local velocity calculated by the desingularized Biot-Savart law \citep{Krasny1986, Zhu2002, Moored2018, Akoz2021}. 
Therefore, a matrix representation of the boundary condition is formed and solved for the body doublet strengths at each time step. 
The perturbation velocity over the surface of the body is then determined by a local differentiation of the perturbation potential.

Following the work of \cite{Moored2018}, the pressure exerted on the body is calculated by the unsteady Bernoulli equation as
\begin{align} \label{eq:unsteadybernoulli} 
P(x,z,t)=-\rho \frac{\partial \Phi^*}{\partial t} + \rho (\mathbf{u_{rel}} + \mathbf{U_{\infty}}) \cdot \nabla \Phi^* -\rho \frac{(\nabla \Phi^*)^2}{2},
\end{align}
where $\mathbf{u_{rel}}$ is the local surface velocity of the pitching foil relative to the body-fixed frame of reference.  The inviscid instantaneous resultant force is then simply an integration of the instantaneous pressure field acting over the body surface. Finally, the resultant force is decomposed into its lift and drag directions.  For more details see \citet{Moored2018}.


\subsection{Force decomposition approach}\label{sec:decomp}
Following the classic force decomposition of \citet{Sears1938} and \citet{Mccune2014}, the lift force can be decomposed as the sum of the added mass lift, quasi-steady lift, and wake-induced lift, which are denoted by $(\cdot)_{a}$, $(\cdot)_{q}$, and $(\cdot)_{w}$,  respectively. We apply this decomposition approach directly to the pressure field defined by the unsteady Bernoulli equation.  

To calculate the added mass pressure, $P_a$, potential flow simulations are employed where there are no wake elements shed and no Kutta condition enforced. Under these conditions the first term in Equation (\ref{eq:unsteadybernoulli}) is known as the added mass pressure. To calculate the quasi-steady pressure, $P_q$, potential flow simulations are employed where there are no wake elements shed, but there is a Kutta condition enforced (modeled in the boundary element method with a trailing-edge panel that extends far from the foil, effectively to infinity). Then the combined second and third terms in Equation (\ref{eq:unsteadybernoulli}) provide the quasi-steady pressure. The combined added mass and quasi-steady pressures are,
\begin{align} \label{eq:unsteadybernoulli_added} 
P_a(x,z,t) + P_q(x,z,t)=\left[\underbrace{-\rho \frac{\partial \Phi^*}{\partial t}}_{\text{added mass}} + \underbrace{\rho (\mathbf{u_{rel}} + \mathbf{U_{\infty}}) \cdot \nabla \Phi^* -\rho \frac{(\nabla \Phi^*)^2}{2}}_{\text{quasi-steady}}\right]_{\text{no wake}}.
\end{align}
The wake-induced pressure, $P_w$, is calculated as the difference between the total pressure, $P$ (calculated from the full simulations described in Section \ref{sec:num_methods}), and the combined added mass and quasi-steady pressures,
\begin{align} \label{eq:unsteadybernoulli_wake} 
P_w(x,z,t)=P(x,z,t) - \left[P_a(x,z,t) + P_q(x,z,t)\right].
\end{align}
Finally, each force component is acquired by integrating its corresponding pressure field over the body. Then the added mass, quasi-steady and wake-induced lift coefficients can be defined and they sum to the total lift coefficient,
\begin{align}
C_{L} = C_{L_{a}} + C_{L_{q}} + C_{L_{w}}.
\end{align}

\subsection{Force decomposition validation}
\begin{figure}
    \centering
    \includegraphics[width=\textwidth]{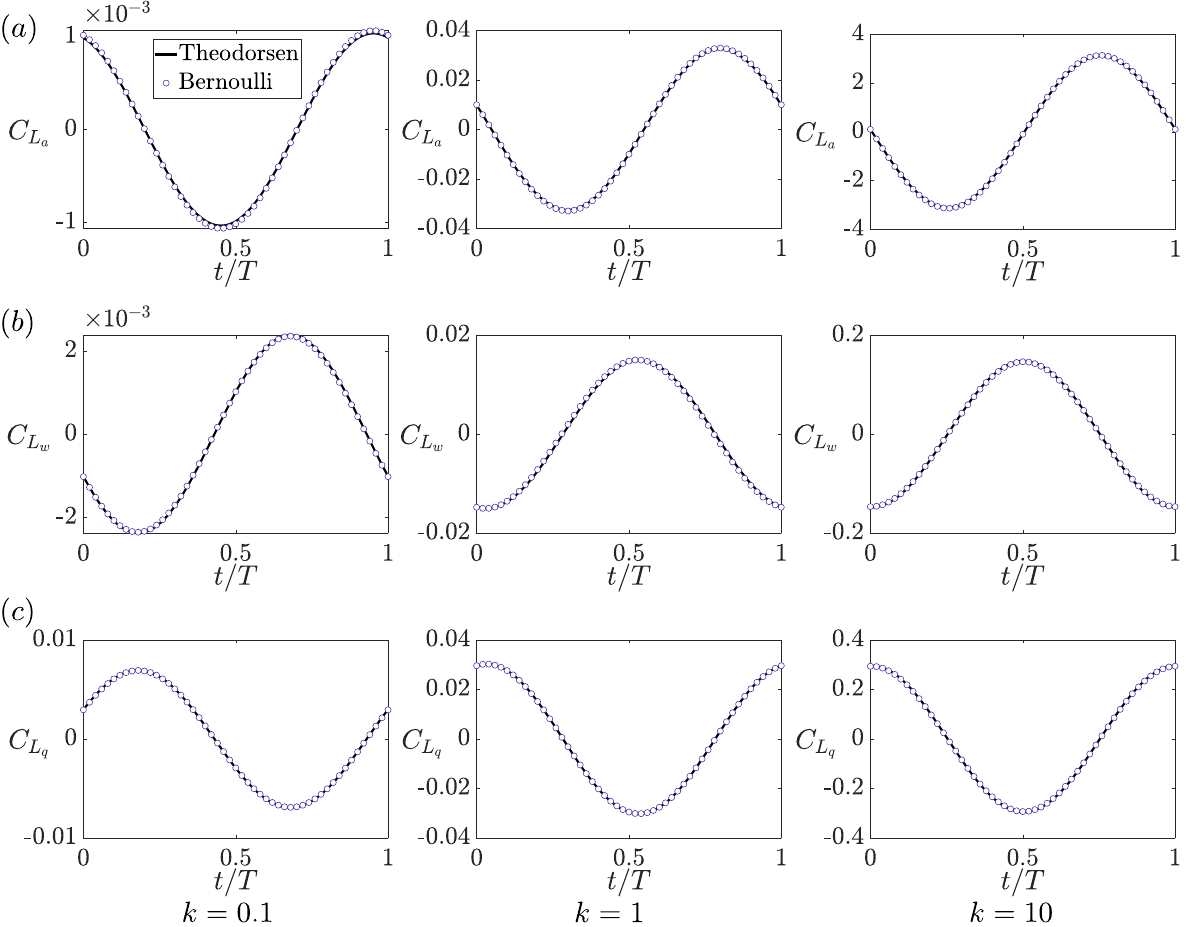}\vspace{0.in}
    \vspace{-0.1in}
    \caption{Validation of the force decomposition approach against Theodorsen's  solution across a wide reduced frequency range of $0.1 \leq k \leq 10$. The subfigure rows show: $(a)$ added mass lift, $(b)$ wake-induced  lift, and $(c)$ quasi-steady lift.} 
   \label{fig:validation}
\end{figure}
Before applying the force decomposition approach to simulations of a pitching foil in ground effect, we first validated it against Theodorsen's solution, which provides a test case where the added mass, quasi-steady, and wake-induced terms are readily identified \citep{bisplinghoff2013aeroelasticity}. In \cite{theodorsen1936}, the lift force generated by a foil pitching about its leading edge is,
\begin{align} \label{Theodorsen} 
L=\underbrace{\rho \pi \frac{c^2}{4}\left(U_{\infty} \dot{\theta} + \frac{c}{2} \ddot{\theta}\right)}_{\text{added mass}} + \underbrace{\rho \pi U_{\infty} c \left(U_{\infty}\theta + \frac{3c}{4}\dot{\theta}\right)}_{\text{quasi-steady}} + \underbrace{\left[C(k)-1\right]\rho \pi U_{\infty} c \left(U_{\infty}\theta + \frac{3c}{4}\dot{\theta}\right)}_{\text{wake-induced}},
\end{align}
where $C(k)$ is the lift deficiency function.
Full boundary element simulations and their decomposition counterparts were performed as described in Sections \ref{sec:num_methods} and \ref{sec:decomp} on a thin airfoil ($1\,\%$ thick tear-drop shape), undergoing small pitching amplitudes of motion about its leading edge ($A^* =0.002$), with a frozen wake (no wake roll-up algorithm applied), and across a wide reduced frequency range of $k = 0.1,\, 1,\,$ and $10$.  Figure \ref{fig:validation} shows that all three lift components precisely match the Theodorsen solution across a wide range of $k$. Even though Theodorsen's solution makes several assumptions, it is worth noting that the force decomposition approach itself is generic and has no underlying assumptions.



\subsection{Experimental methods}
\begin{figure}
    \centering
    \includegraphics[width=0.6\textwidth]{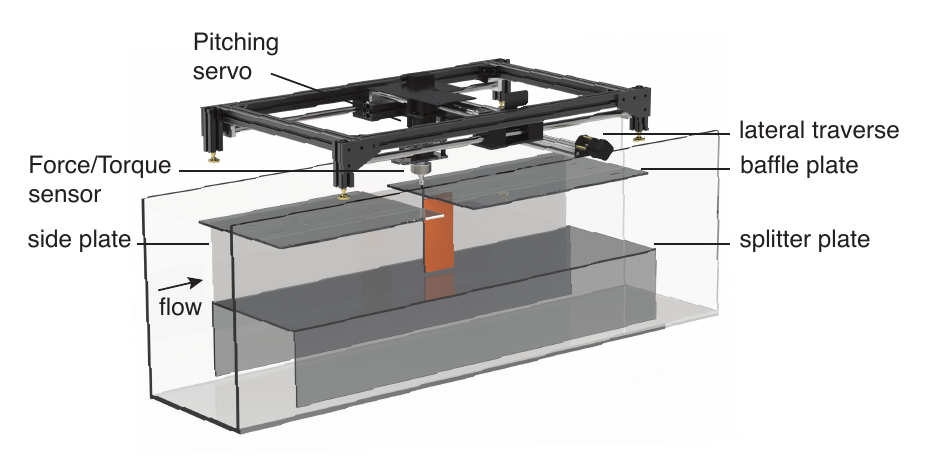}\vspace{0.in}
    \vspace{-0.1in}
    \caption{Schematic of experiment setup equipped with removable splitter plate.} 
   \label{fig:exp setup}
\end{figure}

We conducted four groups of experiments to facilitate the force decomposition analysis, including a 2D static ground effect test, a 2D unsteady high-frequency test (for wake deflection analysis), a 2D and a 3D unsteady ground effect test. All hydrofoils had a tear-drop cross-section with a chord length of $0.095$ m to ensure consistency with simulations, and were 3D-printed with solid ABS. A stainless steel driveshaft with a diameter of  $6.35$ mm was used to pitch the hydrofoil at its leading edge, with the pitch angle $\theta$ being prescribed as $\theta_0 \sin(2\pi f t)$. The driveshaft was driven by a high-torque digital servo motor (Dynamixel MX64), while an absolute encoder (US Digital A2K 4096 CPR) was utilized to verify the angle of the actual motion. The forces and moments generated were measured using a six-axis force-torque sensor (ATI-Mini 40: SI-40-2). The lateral forces were time-averaged over 30 seconds or 20 oscillation cycles to determine the mean lift force for static and unsteady tests. The incoming flow speed was set to 143 $\,$mm/s using an ultrasonic flowmeter (Dynasonics Series TFXB), resulting a chord-based Reynolds number of $13$,$500$. For ground effect tests, an additional splitter plate was used instead of a tunnel sidewall to control the boundary layer thickness (Figure \ref{fig:exp setup}). A boundary layer thickness of $\delta_\mathrm{99\%} \approx 7.5$ mm ($\delta_\mathrm{99\%}/c = 0.08$) was measured using Particle Image Velocimetry at the position aligned with the leading edge of the hydrofoil. 

For all two-dimensional tests, we used a hydrofoil of aspect ratio 3 in a nominally two-dimensional flow condition, which was achieved by installing a horizontal splitter plate and a surface plate near the tips of the hydrofoil. The gap between the hydrofoil tips and the surface/splitter plate was less than 5 mm and surface waves were also minimized by the presence of the surface plate. 

To validate the 3D force decomposition, we repurposed the 3D unsteady ground effect data that was previously published in \cite{Zhong2019}. We tested hydrofoils with four aspect ratios: 1, 1.5, 2 and 2.5, and assessed their performance over a range of ground proximities. Each trial was repeated three times.


\section{Force decomposition results}\label{sec:results}
\begin{table}
\centering
\begin{tabular}{cccccccccccccccccccccccccccccc}
& & & & & Varying $k$  & & & & & & & Varying $St$\\
$k$ & & & & & $0.55\,-\,1$ & & & & & & & $1$\\
$St$ & & & & & $0.3$ & & & & & & & $0.25\,-\,0.55$\\
$D^*$ & & & & & $0.25\,-\,1.5$ & & & & & & & $0.25\,-\,1.5$\\
\end{tabular}
\caption{\label{table:parameter} Simulation variables investigated in the current study.}
\end{table}

A pitching foil in ground effect and constrained to fixed ground distances was simulated across a wide range of dimensionless ground proximities, reduced frequencies, and Strouhal numbers, and force decomposition was applied as described in Section \ref{sec:decomp}.  The simulation variables are detailed in Table \ref{table:parameter}. 

\subsection{Mechanism behind the generation of stable and unstable equilibria}\label{sec:mechanism}
\begin{figure}
    \centering
    \includegraphics[width=\textwidth]{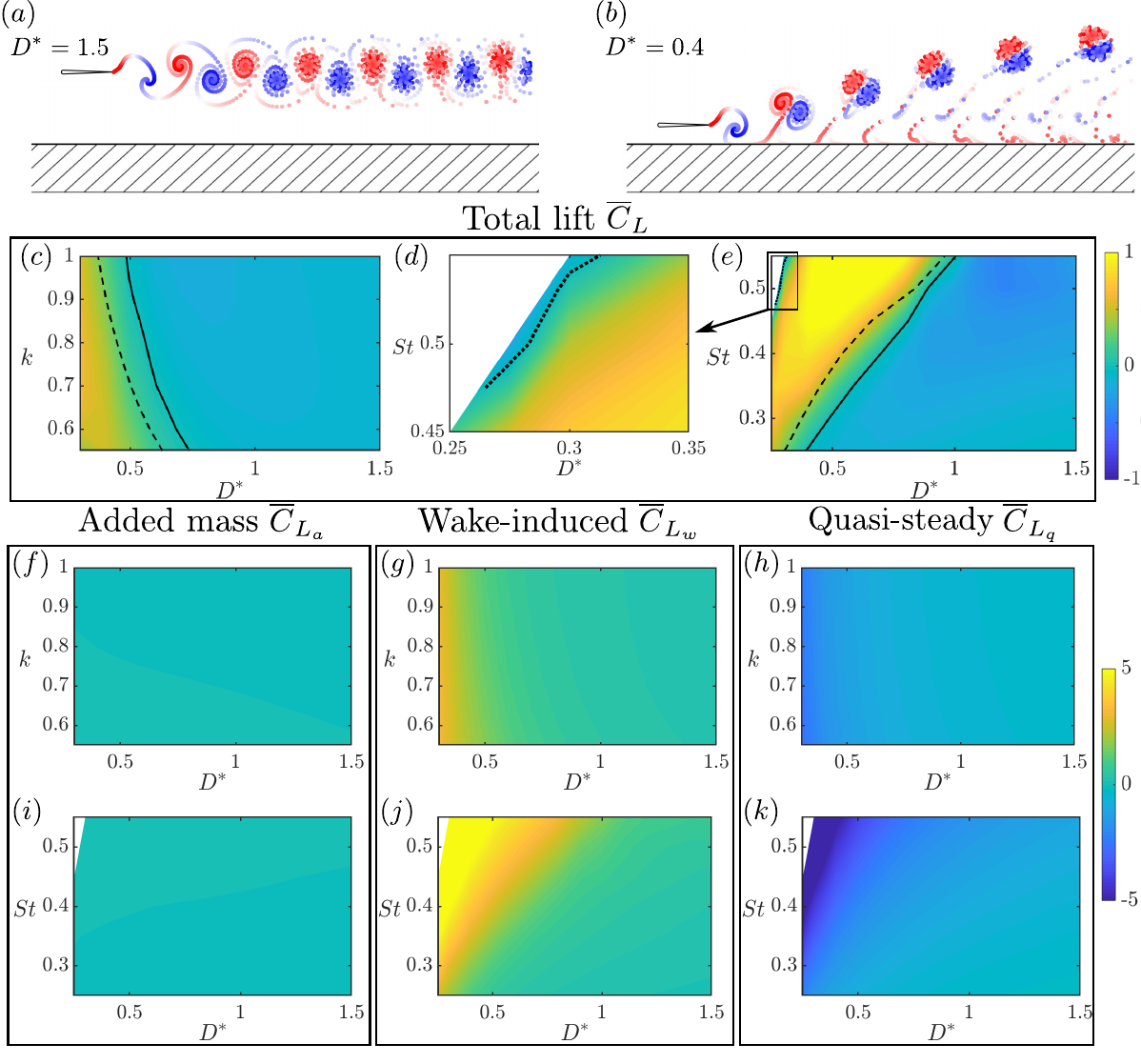}\vspace{0.in}
    \vspace{-0.1in}
    \caption{Both the stable and unstable equilibrium altitudes are generated by a balance between the positive wake-induced lift and negative quasi-steady lift. $(a)$ Wake plots for an out-of-ground effect foil. $(b)$  Wake plots for a near-ground foil. $(c)$ Total lift for varying $k$ with $St =0.3$. $(e)$ Total lift for varying $St$ with $k = 1$. $(d)$ Zoom-in plot highlighting the unstable equilibrium altitude. $(f)$ and $(i)$ added mass lift. $(g)$ and $(j)$ wake-induced lift . $(h)$ and $(k)$ quasi-steady lift. Note that the stable equilibrium altitude is marked by a solid line, the stable equilibrium altitude for a foil unconstrained in the cross-stream direction (mass of the foil is chosen to be $m=2.68\rho s c^2$) is marked by the dashed line, and the unstable equilibrium altitude is marked by the dotted line.
    } 
   \label{fig:totaldecompose}
\end{figure}

\begin{figure}
    \centering
    \includegraphics[width=\textwidth]{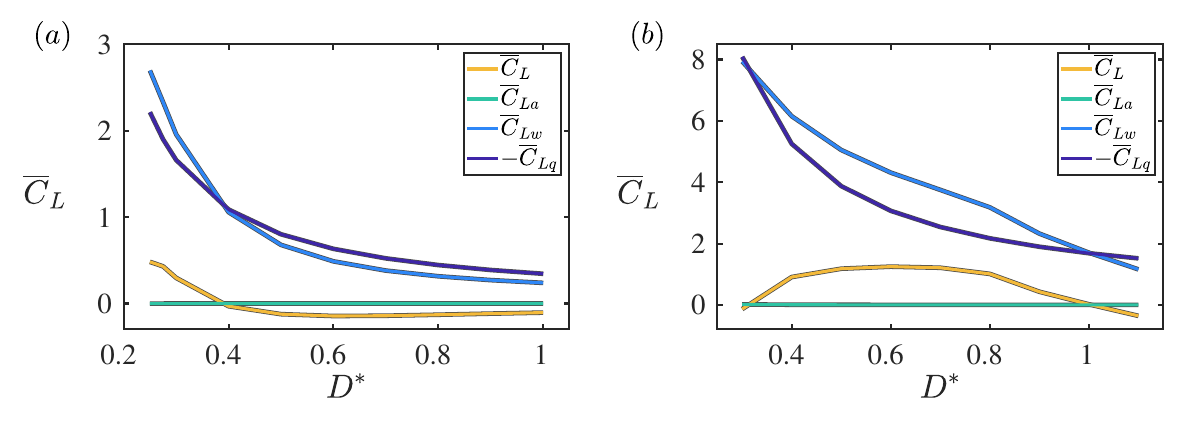}\vspace{0.in}
    \vspace{-0.1in}
    \caption{Force decomposition for $k = 1$: (a) $St = 0.25$ and (b)  $St=0.55$.} 
   \label{fig:detail}
\end{figure}
The pitching foil examined in this study is effectively out of ground effect by $D^* = 1.5$ for cases with $St<0.45$, and by $D^* = 2$ for cases with higher $St$.  In fact, a typical reverse \vk vortex street, indicative of thrust production in an infinite domain, is observed when $D^* = 1.5$, $k = 1$, and $St = 0.3$  (Figure \ref{fig:totaldecompose}a). As the foil approaches the ground to a distance of $D^* = 0.4$ ($k = 1$ and $St = 0.3$), the wake vortices deflect away from the ground (Figure \ref{fig:totaldecompose}b), an effect that has been attributed to the image vortex system \citep{Quinn2014c}.

The total lift for all of the simulations is presented for varying reduced frequency and ground distance ($St = 0.3$; Figure \ref{fig:totaldecompose}c), and for varying Strouhal number and ground distance ($k = 1$; Figure \ref{fig:totaldecompose}d and \ref{fig:totaldecompose}e). A common trend emerges in all of the in-ground effect data except when $St > 0.475$: the lift is positive close to the ground (pushing a foil away from the ground) and negative far from the ground (pulling a foil towards the ground), crossing zero (stable equilibrium altitude; solid lines in Figure \ref{fig:totaldecompose}) between these two regimes. As observed previously, the stable equilibrium altitude increases with increasing $St$ and decreasing $k$ \citep{Kurt2019}, and an unstable equilibrium altitude emerges when $St > 0.475$ (dotted line in Figure \ref{fig:totaldecompose}d), which increases slightly with increasing $St$ in a narrow near-ground region \citep{Zhong2019}. Altitudes below/above the unstable equilibria produce a negative/positive lift, demonstrating that it is indeed unstable. For the first time, this unstable equilibrium altitude is observed in simulation, which corroborates previous experimental findings \citep{Zhong2019}.

The dashed lines in Figure \ref{fig:totaldecompose}c and \ref{fig:totaldecompose}e show the unconstrained simulation results from \citet{Kurt2019}. As discovered in \citet{Cochran-carney2018}, the observed difference in the equilibrium altitudes of the constrained and unconstrained cases is ultimately caused by a difference in the unconstrained simulation's dimensionless mass ($m^* \equiv m/(\rho s c^2) = 2.68$) compared to that of the constrained simulations ($m^* = \infty$).  The finite mass foils of the unconstrained simulations have a finite heaving recoil while the effectively infinite mass foils of the constrained simulations have zero heaving recoil, thereby altering the force production between the two cases and, consequently, their equilibrium altitudes.  Despite these differences, the constrained equilibria show all of the same trends as the unconstrained equilibria with only a small shift in their $D^*$ location.  Moreover, this study will focus on constrained simulations \textit{by design} since these allow us to investigate the forces \textit{outside of equilibrium conditions}, which is essential to fully understand the stability properties of the equilibria.

The decomposition of the lift coefficient into its added mass, wake-induced, and quasi-steady components (Figure \ref{fig:totaldecompose}f--\ref{fig:totaldecompose}k) show clear trends in the components for all ground proximities and kinematics.  The added mass lift is always precisely zero and therefore does not play a role in generating equilibrium altitudes (Figure \ref{fig:totaldecompose}f and \ref{fig:totaldecompose}i).  For in-ground effect data, the wake-induced lift is always positive (Figure \ref{fig:totaldecompose}g and \ref{fig:totaldecompose}j) while the quasi-steady lift is always negative (Figure \ref{fig:totaldecompose}h and \ref{fig:totaldecompose}k), revealing that both unstable and stable equilibria are generated by a balance between positive wake-induced lift and negative quasi-steady lift. The magnitudes of the wake-induced and quasi-steady lift increase with increasing $St$ and decreasing $k$.

The data contours in Figure \ref{fig:totaldecompose} reveal the basic mechanism behind the existence of equilibria, but they do not provide a clear picture as to why there can be both a stable and an unstable equilibria for a given $St$ and $k$.  To better understand this feature of the data the lowest and highest $St$ cases with $k = 1$, $St = 0.25$ and $St = 0.55$, respectively, are presented in Figure \ref{fig:detail}. For low $St$ (Figure \ref{fig:detail}a), the total lift only crosses zero at one $D^*$ producing a stable equilibrium.  This equilibrium altitude is stable since for altitudes above equilibrium the magnitude of the quasi-steady lift is greater than the wake-induced lift leading to a net negative total lift force, and for altitudes below equilibrium the magnitude of the wake-induced lift is greater than the quasi-steady lift leading to net positive total lift. For high $St$ (Figure  \ref{fig:detail}b), the total lift is seen to cross through two equilibrium altitudes.  Now, it becomes clear that the wake-induced lift magnitude curve crosses through the quasi-steady lift magnitude curve at two ground distances because the wake-induced lift has a nearly linear increase with decreasing ground distance while the quasi-steady lift has an exponential growth.  In fact, the slopes of the lift components at the equilibrium points determines their stability: when $\left|d\overline{C}_{Lq}/dD^*\right| < \left|d\overline{C}_{Lw}/dD^*\right|$ the equilibrium altitude is stable and when $\left|d\overline{C}_{Lq}/dD^*\right| > \left|d\overline{C}_{Lw}/dD^*\right|$ the equilibrium altitude is unstable. 

Though this analysis has revealed the basic mechanism behind the generation of stable and unstable equilibrium altitudes, it also raises several questions. How does added mass lift integrate precisely to zero when we expect peak negative added mass lift to be larger in magnitude than the peak positive added mass lift since the foil will get closer and further from the ground throughout an oscillation cycle? Why is the wake-induced lift positive for a wake deflected away from a ground plane when, based on momentum flux arguments, we would expect the opposite? Why is the quasi-steady lift negative in nature? To address these questions, the following sections will examine each force component more deeply.

\subsection{Added mass lift}\label{sec:addedmass}
\begin{figure}
    \centering
    \includegraphics[width=0.8\textwidth]{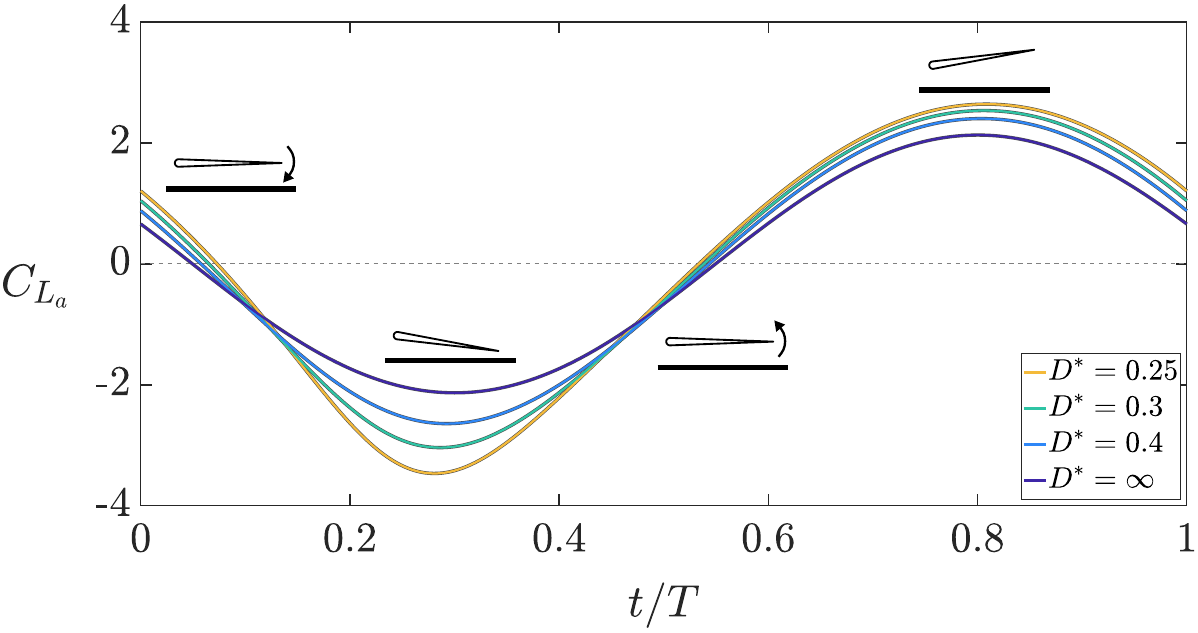}\vspace{0.in}
    \vspace{-0.1in}
    \caption{The added mass lift as a function of time for varying ground distance with $k = 1$ and $St = 0.3$.  Out-of-ground effect data is represented by $D^* = \infty$.}
   \label{fig:addedprofile}
\end{figure}

As shown in the previous section, the time-averaged added mass lift is precisely zero for all of the simulation cases.  This is expected for solutions to the linear potential flow equations with harmonic boundary conditions -- a `scallop theorem' result \citep{purcell1977}. However, for the in-ground effect data it is well-known that as the ground distance decreases there is an increase in the added mass of a foil \citep{Brenner1982,Mivehchi2021}.  This effect should amplify both the positive and negative peaks in the added mass lift, but should more greatly amplify the negative peak, which occurs at the bottom of the down stroke when the foil is closest to the ground.  Considering only these peak forces, one would imagine that the time-averaged added mass lift would be negative, yet we know that it is zero from the data in Section \ref{sec:mechanism}, so how does the added mass lift integrate precisely to zero and satisfy a scallop theorem? 

Figure \ref{fig:addedprofile} presents the time-varying added mass lift at four ground distances when $k = 1$ and $St = 0.3$.  Indeed, as the ground distance decreases both the positive and negative peaks grow in magnitude, and it is true that the negative peak is amplified more than the positive peak creating a net negative lift in the peak forces.  However, it is now clear that the negative peak becomes sharper while the positive peak becomes broader.  When integrating the lift over a full cycle, this effect counteracts the net imbalance in the positive and negative peak forces, leading to a precisely zero time-averaged force.

\subsection{Wake-induced lift}
\begin{figure}
    \centering
    \includegraphics[width=\textwidth]{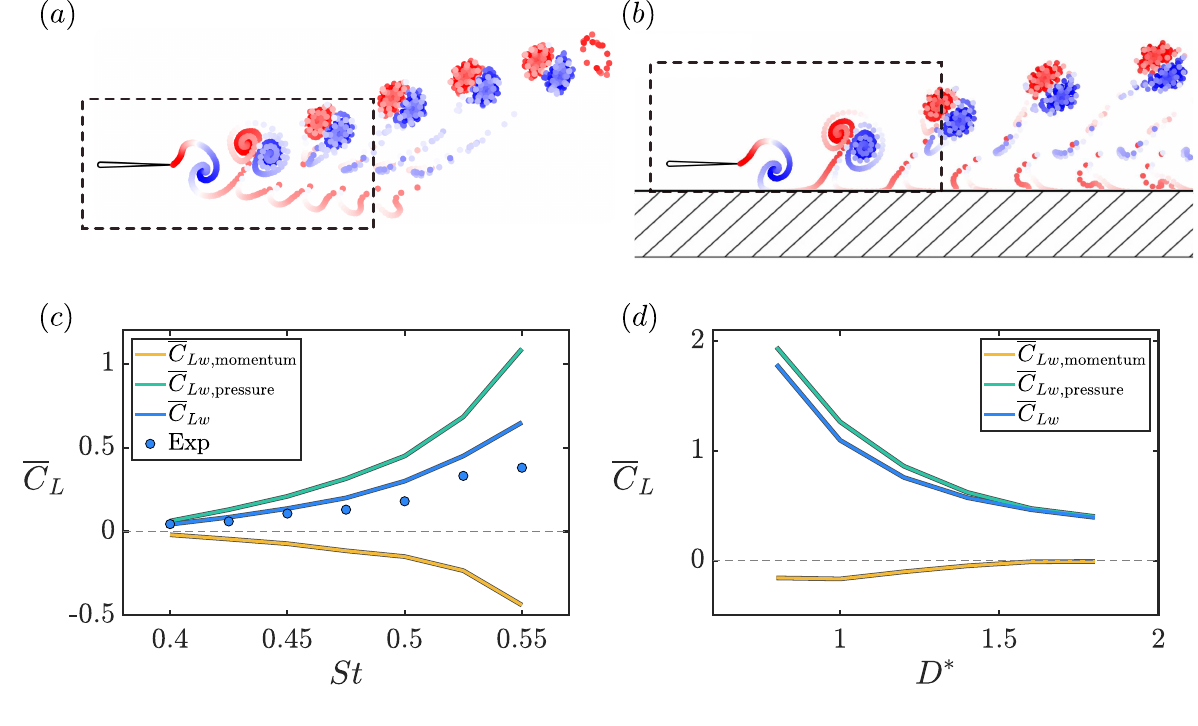}\vspace{0.in}
    \vspace{-0.1in}
    \caption{Time-averaged lift control volume analysis. The control volume is represented by a dashed box. (a) Upward deflected wake generated by an out-of-ground effect foil pitching at $St = 0.55$ and $A^*=0.3$. (b) Upward deflected wake generated by an in-ground effect foil pitching at $St = 0.3$ and $A^*=0.3$. (c) Control volume analysis for the out-of-ground effect case. (d) Control volume analysis for the in-ground effect case. The time-averaged wake-induced lift coefficient from simulations and experiments are denoted by the solid blue lines and blue circles, respectively. Contributions to simulations' wake-induced lift from the pressure difference term and the momentum flux term in the control volume analysis are shown as green and yellow solid lines, respectively.} 
   \label{fig:wakeins}
\end{figure}
In Section \ref{sec:mechanism}, the wake-induced lift was found to be positive for in-ground effect foils regardless of their ground distance and kinematics. This is surprising since the wake vortices deflect away from the ground and in the time-average they also produce a momentum jet deflected away from the ground \citep{Quinn2014c}.  If one imagines a control volume around a foil and considers only the momentum flux term, the deflection of the jet away from the ground should have an associated negative lift force, that is, a force pushing the swimmer toward the ground \citep{Kurt2019}, and yet the data from Section \ref{sec:mechanism} shows otherwise.

To understand this phenomenon more deeply we employ a control volume analysis. To begin we will consider a simple out-of-ground effect case where a deflected wake is also generated very much akin to the deflected wake of in-ground effect foils.  This case is useful since it will help unambiguously show the connection among the wake-induced lift, the pressure field, and the momentum flux, and it presents an opportunity to experimentally verify the trends in the wake-induced lift. With this basic knowledge in mind, we will then proceed to the in-ground effect case.

When a two-dimensional pitching foil oscillates at sufficiently high Strouhal number, it is well-documented \citep{marais2012stabilizing,Das2016} that there is a wake instability that causes the wake vortices to form vortex dipoles and, in turn, deflect the wake. Figure \ref{fig:wakeins}a shows an example of this where a pitching foil is operated at $St = 0.55$ and $A^* = 0.3$.  Since this pitching foil is out of ground effect and has a symmetric motion, then the time-averaged added mass and quasi-steady lift must both necessarily be zero.  However, due to the wake deflection the time-averaged wake-induced lift will be non-zero and, in fact, it will be the only component thus making it equivalent to the total lift in this specific case. 

Following the work of \cite{Cleaver2012}, a time-averaged control volume analysis can be applied to the control volume (dashed box) defined in Figure \ref{fig:wakeins}a, 
\begin{align} \label{eq:controlvolume} 
\overline{L}_{w}=\underbrace{\overline{\int_{UL} \Delta P_{UL}dA}}_{\text{Pressure difference}} \underbrace{-\overline{\int_{CS} \rho v(\mathbf{V}\cdot\mathbf{n})dA}}_{\text{Momentum flux}},
\end{align}
where the total time-averaged lift is precisely equal to the time-averaged wake-induced lift, $\overline{L}_{w}$, the pressure difference between the upper and lower ($UL$) control surfaces is $\Delta P_{UL}$, the velocity field is $\mathbf{V}$, the vector normal to the control surfaces ($CS$) is $\mathbf{n}$, and the horizontal and vertical components of the velocity field are $u$ and $v$, respectively. 

Just like the in-ground effect data from Section \ref{sec:mechanism}, both simulations and experiments show that the wake-induced lift is in fact positive for an upward deflected wake (Figure \ref{fig:wakeins}c) which, in the case of this out-of-ground effect foil, is consistent with previous findings \citep{Emblemsvag2002,Liang2011,Yu2012a,Cleaver2012}. Using the simulation data, Figure \ref{fig:wakeins}c further shows that the wake-induced lift from the net momentum flux term is indeed negative, as expected, while the contribution from the pressure difference term is positive and, surprisingly, is larger in magnitude than the momentum flux term, thereby making the wake-induced lift positive for an upward deflected wake. In light of this analysis, momentum flux considerations alone do not accurately discern the direction of forces generated by an asymmetric wake; the pressure field must also be considered.

Moreover, the same control volume analysis was applied to an in-ground effect pitching foil in Figure \ref{fig:wakeins}b. Since the time-averaged added mass lift is zero (Section \ref{sec:addedmass}), the time-averaged wake-induced lift from the pressure difference and  momentum flux terms are acquired by calculating those two contributions for the total and quasi-steady pressure and flowfields and taking their differences as,
\begin{gather} 
\overline{L}_{w,\text{pressure}}= \overline{\int_{UL} \Delta P_{UL}dA}\bigg|_{\text{total}} - \overline{\int_{UL} \Delta P_{UL}dA}\bigg|_{\text{quasi-steady}} \\
\text{and} \;\; \overline{L}_{w,\text{momentum}} = -\overline{\int_{CS} \rho v(\mathbf{V}\cdot\mathbf{n})dA}\bigg|_{\text{total}} + \overline{\int_{CS} \rho v(\mathbf{V}\cdot\mathbf{n})dA}\bigg|_{\text{quasi-steady}}.
\end{gather}
Figure \ref{fig:wakeins}d shows that in ground effect, the momentum flux term is negative, yet the wake-induced lift acts in the positive direction since the pressure difference term again outweighs the momentum flux term, as in the out-of-ground effect case. It is worth noting that compared to the out-of-ground effect case the pressure difference term now has a more significant effect on the wake-induced lift.

\subsection{Quasi-steady lift}\label{sec:quasi}
\begin{figure}
    \centering
    \includegraphics[width=\textwidth]{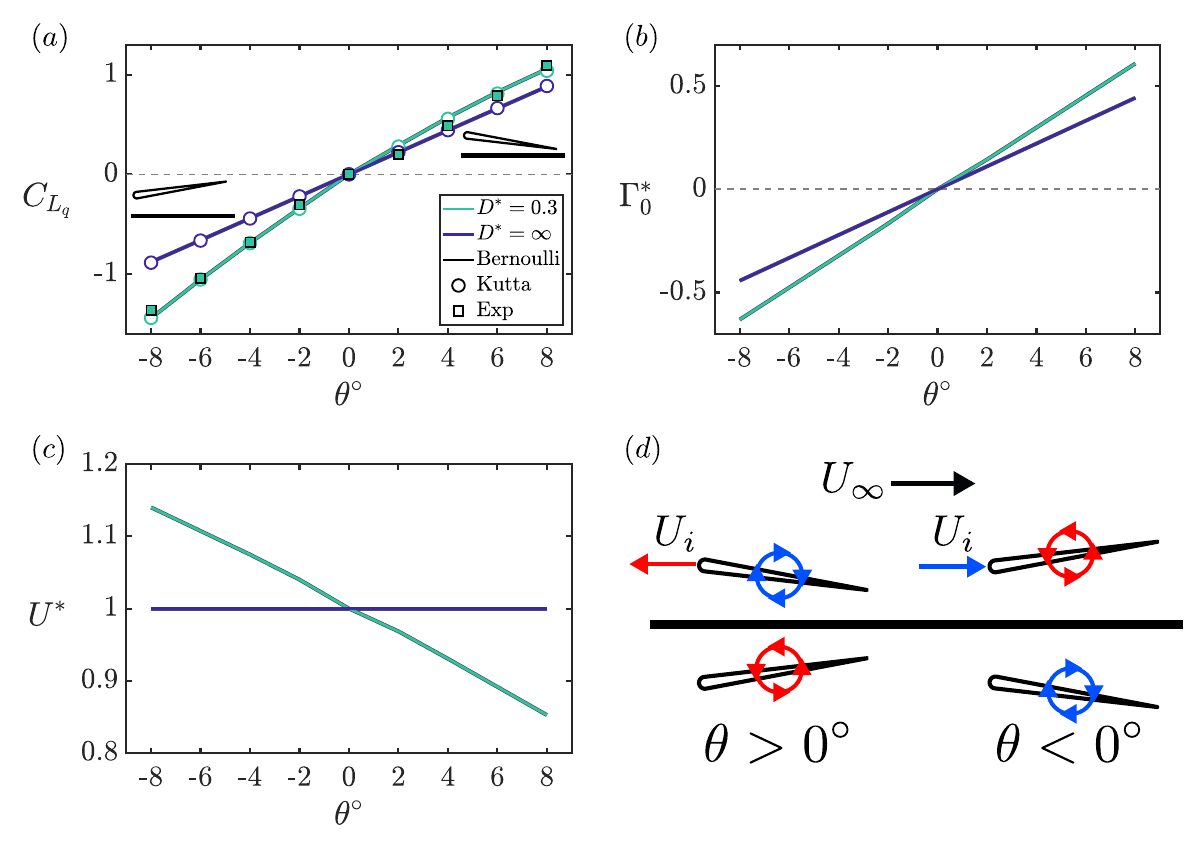}\vspace{0.in}
    \vspace{-0.1in}
    \caption{Asymmetry in the quasi-steady lift for a steady foil in ground effect is caused by the velocity induced on a foil by its image body's quasi-steady bound circulation. $(a)$ The quasi-steady lift coefficient. Note that the lift coefficient calculated in the simulations (using the unsteady Bernoulli equation) is represented with solid lines, the lift coefficient calculated from the simulations using the Kutta-Joukowski theorem is represented by the circle markers and the experimental data is represented by the square markers. $(b)$ The dimensionless quasi-steady circulation. $(c)$ The dimensionless effective flow velocity. $(d)$ Schematic of the components of the effective flow velocity, which is the summation of $U_{i}$ and $U_{\infty}$. } 
    \label{fig:quasisteadydatastill}
\end{figure}

\begin{figure}
    \centering
    \includegraphics[width=\textwidth]{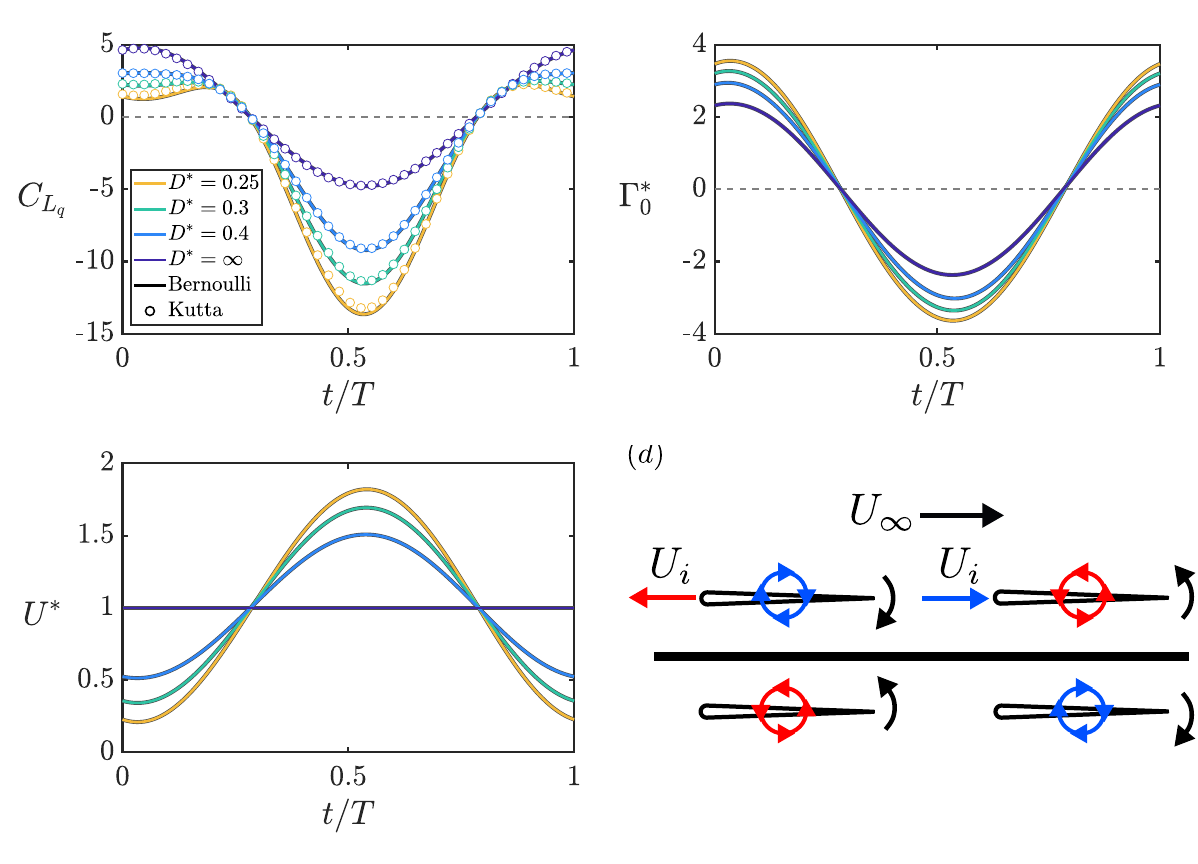}\vspace{0.in}
    \vspace{-0.1in}
    \caption{The velocity induced by the ground image body creates asymmetry in the quasi-steady lift at finite Strouhal numbers just like the steady foil case. $(a)$ The quasi-steady lift coefficient. $(b)$ The dimensionless quasi-steady circulation. $(c)$ The dimensionless effective flow velocity. $(d)$ Schematic showing the effective flow velocity is the summation of $U_{i}$ and $U_{\infty}$.} 
    \label{fig:quasisteadydata}
\end{figure}

  To provide deeper insight as to why there is a net negative quasi-steady force in the time average, we further break down the quasi-steady lift into its constituent components by leveraging the Kutta-Joukowski theorem \citep{Sears1938}, $L_{q} = \rho s U \Gamma_{0}$, where $L_q$ is the quasi-steady lift, $s$ is the span length, $U$ is the local effective flow velocity acting on a foil, and $\Gamma_{0}$ is the quasi-steady bound circulation of a foil.  The local effective velocity is calculated from simulations by summing the freestream velocity, $U_\infty$, with the induced velocity $U_{i}$ at the leading edge from the image body. The quasi-steady bound circulation is simply the negative of the strength of the trailing-edge panel used to enforce the Kutta condition in the boundary element simulations, that is, $\Gamma_0 = -\mu_{TE}$ \cite[]{KatzJoseph;Plotkin2005}. The Kutta-Joukowski theorem can be non-dimensionalized and written as
\begin{align}\label{eq:1}
C_{L_q} = 2 U^* \Gamma_0^* \quad  \; U^* = \frac{U}{U_\infty} \quad \; \Gamma_0^* = \frac{\Gamma_0}{cU_\infty}. 
\end{align}
\noindent When a foil is out of ground effect then $U^* = 1$, $\Gamma_{0}^* = \Gamma_{0,\infty}^*$, and $C_{L_q} = C_{L_{q,\infty}}$, where $\Gamma_{0,\infty}^*$ and $C_{L_{q,\infty}}$ are the time-varying quasi-steady bound circulation and lift coefficient for a foil in an infinite domain. Without loss of generality, the quasi-steady bound circulation for in-ground effect foils can be written as the multiplication of a circulation amplification factor $\beta$ and the infinite domain quasi-steady circulation, $\Gamma_{0}^* = \beta \Gamma_{0,\infty}^* $. We expect that both the circulation and the effective flow velocity will be amplified or reduced for foils in ground effect as opposed to out-of-ground effect foils.  

To understand the connections among the circulation amplification factor, the amplified/reduced effective flow velocity, and the quasi-steady lift, as well as to verify these connections with experimental data, we begin with a simple case of a static foil in and out of ground effect at various pitch angles (Figure \ref{fig:quasisteadydatastill}).  For the static foil, the quasi-steady lift is precisely the total lift, which is readily measured in experiments.  Figure \ref{fig:quasisteadydatastill}a presents the quasi-steady lift for $D^* = 0.3$ and $D^* = \infty$ from the simulations (solid lines), which is directly determined by integrating the pressure acting on the foil calculated from the unsteady Bernoulli equation (Section \ref{sec:num_methods}).  The simulations are also used to calculate the quasi-steady lift by using the Kutta-Joukowski theorem (circle markers) as outlined above, and both approaches are observed to produce exactly the same result.  Also, the in-ground effect quasi-steady lift is verified by having good agreement with experimental data (square markers).  

For the out-of-ground effect foil, the quasi-steady lift shows the classic linear relationship with the pitch angle for $-8^\circ \leq \theta \leq 8^\circ$.  However, when in ground effect the quasi-steady lift shows a nonlinear relationship with the pitch angle, where both the negative and positive lift regimes are amplified compared to the out-of-ground effect case, yet the negative lift regime is amplified more than the positive lift regime.  This produces an asymmetric lift response about $\theta = 0$. If one were to quasi-statically pitch the foil through an $8^\circ$ amplitude sinusoidal pitching motion this asymmetric lift response would lead to a net negative quasi-steady lift in the time-average, just as is observed in the quasi-steady data of finite Strouhal numbers. In fact, throughout this section we will show that precisely the same mechanism is at play in this static case as is in the quasi-steady cases of finite Strouhal number.

Figure \ref{fig:quasisteadydatastill}b presents the dimensionless quasi-steady circulation.  The circulation amplification is observed to be symmetric about $\theta = 0$ and appears to just be a constant factor giving rise to a slope change in the $\Gamma_0^*$ -- $\theta$ line.  This shows, at least for the static case, that $\beta$ is only a function of $D^*$, not $\theta$. Moreover, the circulation amplification is responsible for amplifying the quasi-steady lift forces in ground effect, but is not responsible for asymmetry in the lift curve.  Figure \ref{fig:quasisteadydatastill}c shows that the effective flow velocity increases above the freestream speed when the quasi-steady circulation is negative (counterclockwise),  while it decreases below freestream speed when the quasi-steady circulation is positive (clockwise). This occurs due to the velocity induced by the quasi-steady bound vortex of the image body (Figure \ref{fig:quasisteadydatastill}d). By consequence, increased effective flow speed acts to amplify the negative lift of negative pitch angles (negative bound circulation) thereby acting in concert with the circulation amplification effect, and decreased effective flow speed acts to reduce the positive lift of positive pitch angles (positive bound circulation), thereby counteracting the circulation amplification effect.  Therefore, it is the change in the effective flow speed that leads to the asymmetric lift response. This mechanism will be implicated further as the reason for the net negative quasi-steady lift for oscillating foils from zero to finite Strouhal numbers.

Figure \ref{fig:quasisteadydata} presents the quasi-steady lift coefficient, dimensionless quasi-steady circulation, and dimensionless effective flow velocity for a foil at a range of ground distances with $St = 0.3$. In Figure \ref{fig:quasisteadydata}a, first, the quasi-steady lift calculated by the unsteady Bernoulli equation and by the Kutta-Joukowski theorem still show good agreement. Figure \ref{fig:quasisteadydata}b shows that the negative and positive quasi-steady circulation are symmetrically amplified with decreasing ground distance, precisely as in the static foil case.  Moreover, Figure \ref{fig:quasisteadydata}c shows that the effective flow velocity increases above and decreases below the freestream velocity when the bound circulation is negative and positive, respectively, also mirroring the static foil case. Consequently, the change in effective flow velocity leads to an enhancement in the negative lift  and a reduction in the positive lift thereby generating a net negative quasi-steady lift in the time average over one pitching cycle. Figure \ref{fig:quasisteadydata}d illustrates precisely the same effective flow altering mechanism of the image body's quasi-steady circulation as is in the static foil case.

\subsection{Three-dimensional decomposition}
\begin{figure}
    \centering
    \includegraphics[width=\textwidth]{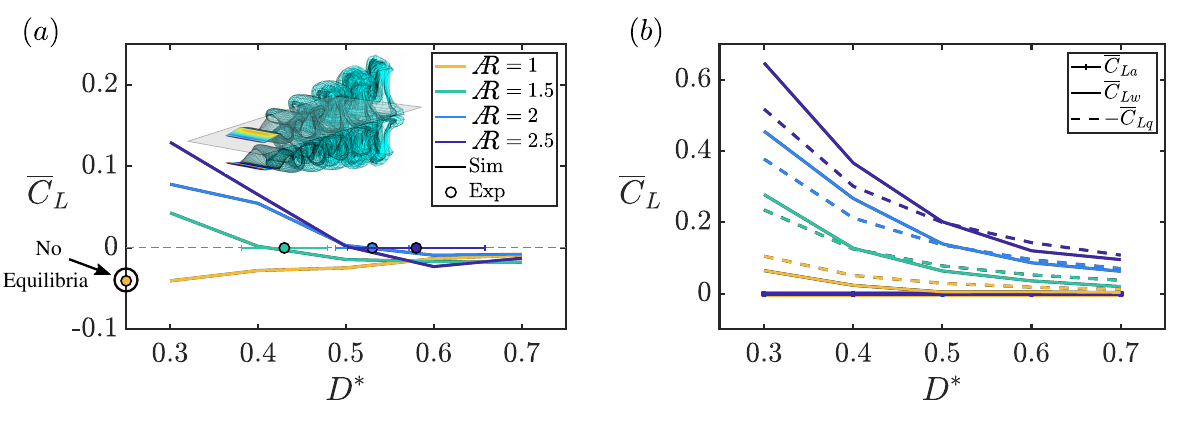}\vspace{0.in}
    \vspace{-0.1in}
    \caption{The magnitude of the wake-induced lift of three-dimensional foils degrades more with decreasing $\AR$ than their quasi-steady lift leading to the disappearance of an equilibrium altitude for $\AR < 1.5$. Simulation data of pitching three-dimensional hydrofoils of varying the aspect ratio with $St= 0.25$ and $\theta_{0} = 11^{\circ}$. $(a)$ Total lift as calculated from the simulations. Note that the circle markers represent the stable equilibrium altitudes measured by experiments. $(b)$ Decomposed lift components from the simulations.   } 
    \label{fig:3Ddecompose}
\end{figure}

In Figure \ref{fig:3Ddecompose}a, both $3D$ simulations and experiments observe that the stable equilibrium altitude disappears when the aspect ratio decreases to $\AR = 1$, which is reproduced from \cite{Zhong2019}. Figure \ref{fig:3Ddecompose}b, presents the lift decomposition of the 3D simulations into the added mass, quasi-steady and wake-induced components. Exactly as in the 2D simulations, a stable equilibrium altitude arises for foils of finite $\AR$ by a balance between the positive wake-induced lift and negative quasi-steady lift. The equilibrium altitude disappears for low aspect ratio ($\AR < 1.5$) hydrofoils, which instead experience a net negative time-averaged lift force. The decomposition reveals that this occurs since the magnitude of the wake-induced lift reduces more with decreasing $\AR$ than the reduction in the magnitude of quasi-steady lift, thereby having the negative quasi-steady lift outweigh the positive wake-induced lift at all ground distances.

\subsection{Scaling the equilibrium altitude}
\begin{figure}
    \centering
    \includegraphics[width=\textwidth]{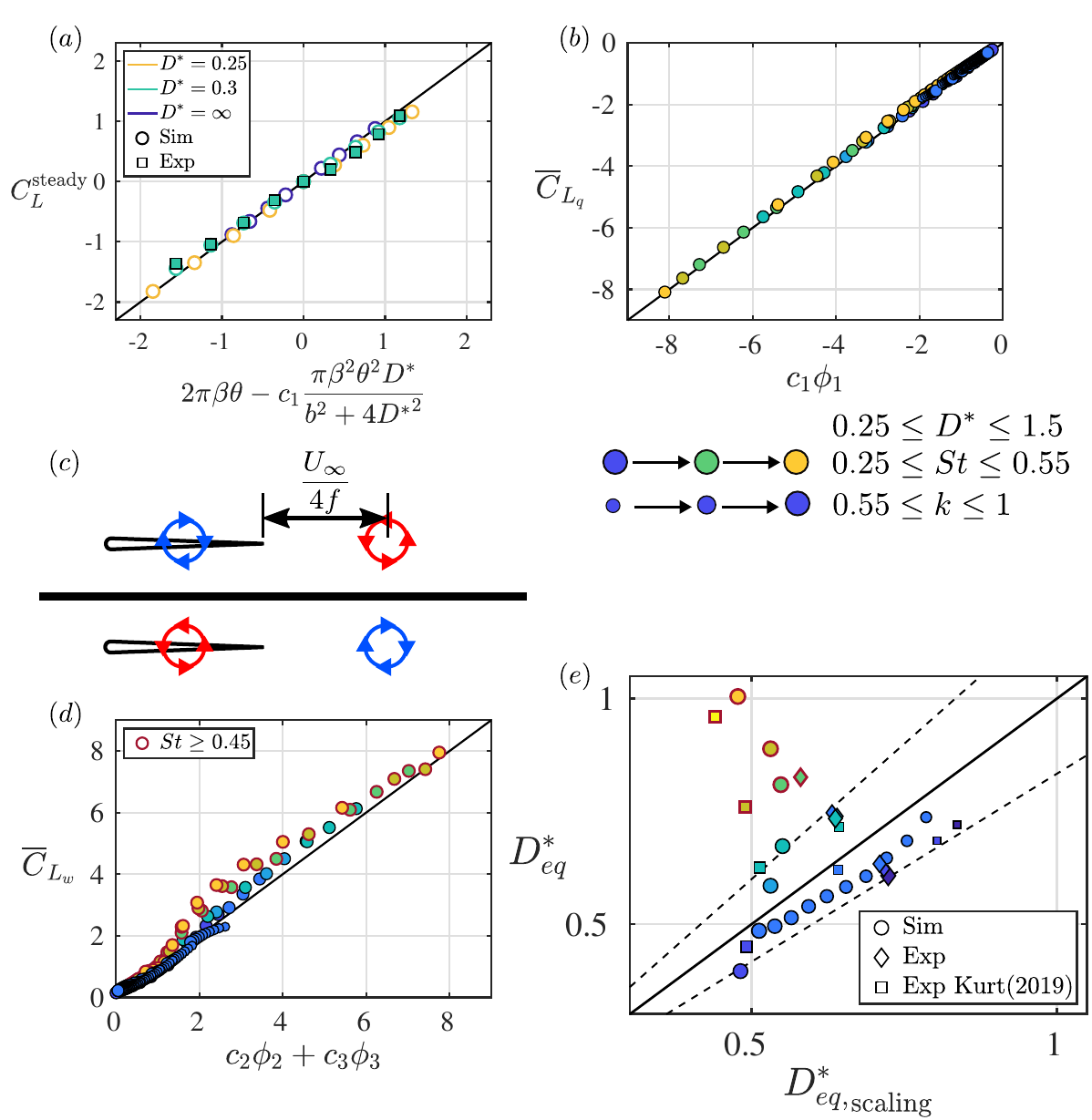}\vspace{0.in}
    \vspace{-0.1in}
    \caption{Scaling laws for steady and unsteady ground effect show a good collapse across a wide range of $St$, $k$, and $D^*$. (a) The lift for a steady foil in ground effect (data from Figure \ref{fig:quasisteadydatastill}a with additional numerical case at $D^*=0.25$) is compared to the steady foil scaling law. $(b)$ The quasi-steady lift of an unsteady pitching foil (data from Figure \ref{fig:quasisteadydata}) compared to its scaling relation for $0.25 \leq D^* \leq 1.5$, $0.25 \leq St \leq 0.55$, and $0.55 \leq  k \leq 1$. The marker color from dark blue to light yellow represents increasing $St$ while the marker size from small to large represents increasing $k$. $(c)$ Schematic to illustrate the simple flow model for calculating the wake circulation amplification factor. $(d)$ The wake-induced lift of an unsteady pitching foil compared to its scaling relation. Markers outlined in red represent data with $St \ge 0.45$.  $(e)$ Stable equilibrium altitudes compared to the scaling relation. Dashed lines present $20\%$ margins of error. These include numerical data with $0.25 \leq St \leq 0.55$ and $0.55 \leq k \leq 1$, and experimental data with $0.2 \leq St \leq 0.4$ and $0.6 \leq k \leq 0.9$, as well as experimental data from \cite{Kurt2019} with $0.2 \leq St \leq 0.6$ and $0.5 \leq k \leq 1$. } 
    \label{fig:scaling}
\end{figure}

The analysis of the added mass, quasi-steady, and wake-induced forces presented above can now be leveraged to develop scaling laws that capture the physics of ground effect and are capable of predicting the equilibrium altitude of a pitching foil. Since the added mass forces do not play a role in the generation of an equilibrium altitude then scaling laws for only the quasi-steady and wake-induced lift are sought.  
\subsubsection{Quasi-steady lift}
As presented in Section \ref{sec:quasi}, the quasi-steady lift coefficient is,
\begin{gather}
C_{L_{q}} \propto  U^* \beta \Gamma_{0,\infty}^* ,
\end{gather}
\begin{gather}
\Gamma_{0,\infty}^* = \pi\theta +\frac{\pi c \dot{\theta}}{2U_{\infty}} \label{eq:2},
\end{gather}
\noindent where $\beta$ is the quasi-steady circulation amplification factor, $U^*$ is the normalized effective flow velocity at the foil's leading edge, and $\Gamma^*_{0,\infty}$ is simply the formula for a flat plate pitching about its leading edge from \cite{theodorsen1936} and \cite{Mccune2014}. To determine a scaling law for the quasi-steady lift, consider a simplified flow model where the quasi-steady bound vortex is located along the chord (Figure \ref{fig:quasisteadydata}d). Then a scaling relation for the circulation amplification factor and the effective flow velocity can be determined. The circulation amplification factor is calculated by finding the circulation amplification necessary to enforce the Kutta condition at the foil's trailing edge due to the influence of the image quasi-steady bound vortex. Assuming that the distance between the bound vortex and foil's leading edge is $bc$ with $0\leq b \leq 1$, then the circulation amplification factor is simply calculated as,
\begin{gather}
\beta = 1 + \frac{(1-b)^2}{4{D^*}^2} \quad \text{where} \quad b = 0.55.
\end{gather}
 To determine the placement of the bound vortex along the chord, circulation amplification data at $D^*=0.5$ from an exact solution for a steady foil in ground effect (Figure 4d in \cite{Baddoo2020}) was used to determine that $b=0.55$. The effective flow velocity is calculated by finding the normalized induced velocity at the leading edge, $U^*_{i} =U_{i}/U_\infty$, from the image foil's quasi-steady bound vortex,
\begin{gather}
 U^* = 1 + U^*_{i} \quad \text{where} \quad U^*_{i} = \frac{-\Gamma_{0,\infty}^*\beta}{\pi}\left(\frac{D^*}{b^2 + 4{D^*}^2}\right).
\end{gather}

\noindent After organization, the time-varying quasi-steady lift becomes,
\begin{gather}
C_{L_{q}} \propto  \beta\Gamma_{0,\infty}^* - \frac{\beta^2{\Gamma^*}^2_{0,\infty}D^*}{\pi(b^2+4{D^*}^2)}\label{eq:3}.
\end{gather}
 According to thin airfoil theory, the coefficient of the first term is $2$, which gives,
 \begin{gather}
C_{L_{q}} = 2   \beta\Gamma_{0,\infty}^*-c_{1}\frac{\beta^2{\Gamma^*}^2_{0,\infty}D^*}{\pi(b^2+4{D^*}^2)},\label{eq:4}
\end{gather}
where $c_{1}$ needs to be determined.
 
 As an important side note, this formula can be used to determine a lift scaling relation for \textit{steady} ground effect by neglecting the rotational lift or virtual camber term  ($\dot{\theta}$ term) in $\Gamma^*_{0,\infty}$. Then a scaling law for steady ground effect can be deduced as,
 \begin{gather}
 C^\text{steady}_L = 2\pi\beta\theta - c_1\frac{\pi\beta^2\theta^2 D^*}{b^2+4{D^*}^2} \quad \text{where} \quad c_{1} = 2.84.
 \end{gather}
\noindent Using the static foil numerical and experimental data from Figure \ref{fig:quasisteadydatastill}a the coefficient is determined to be $c_1 = 2.84$ by linear regression. Figure \ref{fig:scaling}a then compares the scaling law prediction with the data, which shows that it can predict the steady ground effect data to within $10\%$ for the numerical data and $15\%$ for the experimental data.  
 
Returning to the the time-varying formula, Equation (\ref{eq:4}), we can also consider an \textit{unsteady} ground effect case where the foil undergoes sinusoidal pitching. If this scaling relation is then time-averaged only a contribution from the second term will remain as,
 \begin{gather}
\overline{C}_{L_{q}} =  -c_{1}\frac{\beta^2\overline{{\Gamma^*}^2_{0,\infty}}D^*}{\pi(b^2+4{D^*}^2)}.
\end{gather}
Considering that $\theta = \theta_0 \sin (2\pi f t)$ and $\theta_{0} = A^*/2=St/(2k)$ we have,
 \begin{gather}
 \overline{C}_{L_{q}} = c_{1}\phi_{1},
\\ \phi_{1} =   -\frac{\pi^3}{4}\left(\frac{\beta^2 D^*}{b^2+4{D^*}^2}\right) \left(1+ \frac{1}{\pi^2k^2} \right) St^2.
\end{gather}
The coefficient $c_{1}=2.84$ determined from the static foil data is used for scaling the quasi-steady lift of both the steady and unsteady foil cases. Figure \ref{fig:scaling}b presents an excellent collapse of the quasi-steady lift data  over a wide range of $St$, $k$ and $D^*$. 

\subsubsection{Wake-induced lift}
The wake-induced lift can be calculated as,
\begin{gather}
C_{L_{w}} \propto  U^* \gamma \Gamma_{w,\infty}^*,
\\ U^* = 1 + U^*_{i,1} + U^*_{i,2}.
\end{gather}
where $\Gamma_{w,\infty}^*$ is the normalized infinite domain wake circulation, and the effective flow velocity $U^*$ is composed of the velocity $U_{i,1}^*$, induced by an image bound vortex (representing the total bound vorticity, not the quasi-steady bound vorticity), and the velocity induced by an image wake vortex, $U_{i,2}^*$. To determine a scaling law for the wake-induced lift, consider a simplified flow model where a wake vortex is placed along the chord line at a distance of $U_\infty/(4f)$ downstream of the trailing edge (Figure \ref{fig:scaling}c), which gives,
\begin{gather}
U^*_{i,1} = \frac{-\Gamma_{\infty}^*\gamma}{\pi}\left(\frac{D^*}{b^2 + 4{D^*}^2}\right) \;\; \text{and} \;\; U^*_{i,2} =  \frac{-\Gamma_{w,\infty}^*\gamma}{\pi}\left[\frac{16k^2D^*}{(1+4k)^2  + 64k^2{D^*}^2}\right],
\end{gather}
where $\Gamma^*_{\infty}$ is the normalized infinite domain total circulation, and $\gamma$ is the wake circulation amplification factor. By enforcing the Kutta condition at the foil's trailing edge, the wake circulation amplification factor is calculated as,
 \begin{gather}
\gamma = \frac{1+4k(1-b)}{1+4k(1-b) - \frac{(1-b)^2}{(1-b)^2+4{D^*}^2} +  \frac{4k(1-b)}{1+64k^2{D^*}^2}}.  
\end{gather}
Following \cite{Moored2017}, we have,
\begin{gather}
\Gamma_{w,\infty}^* \propto \Gamma_{0,\infty}^* \frac{k^*}{k^*+1} \;\; \text{and} \;\; k^* = \frac{k}{1+ 4St^2},
\\\Gamma_{\infty}^* \propto \Gamma_{0,\infty}^*.
\end{gather}
Similar to the quasi-steady lift, after time-averaging the wake-induced lift only terms associated with  ${\Gamma^*}^2_{0,\infty}$ remain, which gives,
\begin{gather}
\overline{C}_{L_{w}} = c_{2}\phi_{2} + c_{3}\phi_{3},
\\\phi_{2} = -\gamma^2St^2 \left(1+ \frac{1}{\pi^2k^2} \right)\left(\frac{D^*}{b^2+ 4{D^*}^2}\right)\left(\frac{k^*}{k^*+1}\right),
\\\phi_3 = -\gamma^2St^2 \left(1+ \frac{1}{\pi^2k^2} \right)\left[\frac{16k^2D^*}{(1+4k)^2  + 64k^2{D^*}^2}\right]\left(\frac{k^*}{k^*+1}\right)^2.
\end{gather}
By using the wake-induced lift data from Figure \ref{fig:totaldecompose}, the coefficients were determined by linear regression to be $c_{2} = -136.62$ and $c_{3} = 346.27$. Figure \ref{fig:scaling}d shows that a good collapse of the data can be achieved with the exception of data with $St \ge 0.45$ (outlined in red). As shown in Figure \ref{fig:wakeins}a, the wake-induced lift is significantly altered by a wake instability effect that occurs when $St \ge 0.45$ even for out-of-ground effect foils. Since this wake instability effect is not accounted for in the scaling laws, the data points with $St \ge 0.45$ (Figure \ref{fig:scaling}d) begin to show deviation from the scaling law prediction.

\subsubsection{Total lift}
Now, the quasi-steady lift and wake-induced lift can be summed to acquire the total lift, and the total lift becomes a function of $St$, $k$ and $D^*$,
\begin{gather}
\overline{C}_{L} = c_{1}\phi_{1} + c_{2}\phi_{2} + c_{3}\phi_{3}= f \left(St, k, D^*)\right.\label{eq:5}
\end{gather}
Then for a given pair of $St$ and $k$, the equilibrium altitude $D^*_{eq}$ can be determined by setting Equation (\ref{eq:5}) to zero. Figure \ref{fig:scaling}e presents the scaling law prediction for the equilibrium altitude.  The scaling law can predict the equilibrium altitude to within $20\%$ error except for the data with $St \ge 0.45$. The deviation in this prediction at $St \ge 0.45$ is due to neglecting the wake instability effect in the wake-induced lift scaling law. For the numerical data with $St<0.45$ a $5\%$ and $10\%$ error in predicting the quasi-steady  and wake-induced lift, respectively, leads to a $20\%$ error in predicting the equilibrium altitude. This indicates that the equilibrium altitude relies on a delicate balance and is sensitive to small changes in the lift components. This is also observed by \cite{Liu2023} who found that a pitch bias angle equal to $7\%$ of the amplitude can shift the equilibrium altitude of a pitching foil by $28\%$.


\section{Conclusions}
In this article we decomposed the lift of a near-ground pitching hydrofoil into its added mass, wake-induced,  and quasi-steady components, and determined that their time-averaged values are always zero, positive, and negative, respectively, across all ground proximities. This shows that both stable and unstable equilibrium altitudes are generated by a balance between positive wake-induced lift and negative quasi-steady lift while the added mass lift doesn't play a role.

Using both simulations and experiments detailed analyses are provided to illustrate the three lift components' near-ground behavior. In ground effect the negative peak of the added mass lift is amplified more than the positive peak, however, the negative peak also becomes sharper while the positive peak becomes broader compared to a foil out of ground effect. This leads to a scallop theorem result where the added mass lift precisely integrates to zero over one oscillation cycle. Through a control volume analysis, the wake-induced lift is discovered to be positive due to the positive lift from the pressure difference on the control volume outweighing the negative lift from the net momentum flux. This leads to the upwards deflected wake behind a near-ground swimmer being responsible for a positive wake-induced lift. The quasi-steady lift is further analyzed through the lens of the Kutta-Jowkouski theorem. It's observed that the velocity induced by the positive (negative) quasi-steady circulation bounded to the ground image foil enhances (reduces) the effective flow velocity at the foil's leading edge. For a pitching foil, this amplifies the negative quasi-steady lift during the upstroke and reduces the positive quasi-steady lift during the downstroke leading to a net negative quasi-steady lift. Additionally, the lift decomposition was applied to three-dimensional pitching foil simulations of varying aspect ratio. It was demonstrated that equilibrium altitudes of $3D$ hydrofoils are also generated by a balance between positive wake-induced lift and negative quasi-steady lift. Moreover, as the aspect ratio decreases, the magnitude of the wake-induced lift drops off faster than the magnitude of the quasi-steady lift leading to the disappearance of an equilibrium altitude when $\AR< 1.5$.

Using these insights, scaling laws of the quasi-steady lift, wake-induced lift and equilibrium altitudes are developed. Additionally, a simple scaling law for the lift of a steady foil in ground effect is discovered. The scaling laws show good agreement to both numerical and experimental data by predicting the equilibrium altitudes to within $20\%$ of their value as long as $St<0.45$. For $St \ge 0.45$ significant effects from  wake instability, not accounted for in the scaling laws, arise that alter the wake-induced lift. These results not only provide key physical insights and scaling laws for unsteady ground effect, but also for two schooling hydrofoils in a side-by-side formation with an out-of-phase synchronization.

\section{Acknowledgments}
This work was supported by the National Science Foundation under Program Director Dr. R. Joslin in Fluid Dynamics, award number 1921809 as well as by the Office of Naval Research under Program Director Dr. Robert Brizzolara on MURI grant number N00014-22-1-2616.

\section{Declaration of interests}
The authors report no conflict of interest

\FloatBarrier
\bibliography{library.bib}
\bibliographystyle{jfm}

\end{document}